\documentstyle[12pt,epsf]{article}

\advance\voffset by -1.5cm
\advance\hoffset by -1.5cm
\def\be{\begin{equation}}
\def\ee{\end{equation}}

\def\pmb#1{\setbox0=\hbox{#1}
 \kern-.025em\copy0\kern-\wd0
 \kern.05em\copy0\kern-\wd0
 \kern-.025em\raise.0433em\box0 }

\def\3{\ss}
\def\sq{\hbox{\rlap{$\sqcap$}$\sqcup$}}
\def\qed{\ifmmode\sq\else{\unskip\nobreak\hfil
\penalty50\hskip1em\null\nobreak\hfil\sq
\parfillskip=0pt\finalhyphendemerits=0\endgraf}\fi}
\def\half {\frac{1}{2}}

\def\bbbz {{\sf Z\!\!Z}}

\def\ss{\bf S}

\def\C{{\cal C}}

\def\I{{\cal I}}

\begin{document}

\thispagestyle{empty}
\def\thefootnote{\fnsymbol{footnote}}
\begin{flushright}
  hep-th/9811064\\
  CALT-68-2196
 \end{flushright}
\vskip 0.5cm

\begin{center}\LARGE
{\bf Stable Non-BPS Dyons in N=2 SYM}
\end{center}
\vskip 1.0cm
\begin{center}
{\large  Oren Bergman\footnote{E-mail  address: {\tt
bergman@theory.caltech.edu}}}

\vskip 0.5 cm
{\it Department of Physics\\
California Institute of Technology\\
Pasadena, CA 91125}
\end{center}

\vskip 1.5cm

\begin{center}
November 1998
\end{center}

\vskip 1.5cm

\begin{abstract}
As a novel application of string junctions, we provide evidence for the
existence of stable non-BPS dyons with magnetic charge
greater than 1 in (the semiclassical regime of)
$N=2$ SU(2) Super-Yang-Mills theory.
In addition, we find a new curve of marginal stability.
Moduli space is therefore divided into four regions,
each containing a different stable particle spectrum.
\end{abstract}

\vskip 1.5cm 
\begin{center}
\end{center}

\vfill
\setcounter{footnote}{0}
\def\thefootnote{\arabic{footnote}}
\newpage

\renewcommand{\theequation}{\thesection.\arabic{equation}}

\section{Introduction}

\noindent Type IIB string junctions \cite{schwarz},
and more generally string webs\footnote{By string 
junctions we mean a fundamental string
ending on a D-string, or any of its $SL(2,\bbbz)$ 
counterparts. String webs are composed of several string
junctions.} \cite{sen_net},
have proven to be quite useful
in determining the BPS spectra of supersymmetric
field theories, which can be realized as 
world-volume theories on certain branes
[3--10].
The two fundamental properties of string junctions that 
allow us to learn about the spectrum of BPS states
is that their total NSNS and RR charges vanish
\cite{schwarz}, and
that the orientations of the strings are
correlated with their $(p,q)$ type such that the net
force on the junction vanishes \cite{dm}.
For planar webs, the second property guarantees that the configuration 
is (classically) supersymmetric \cite{sen_net}.

Previously we constructed the BPS states of $N=2$
$SU(2)$ SYM using string webs \cite{bf1}. 
The theory is realized as the world-volume field theory 
of a 3-brane probe in the background of two mutually
non-local 7-branes \cite{bds,sen_bps}, and the BPS states 
correspond to supersymmetric string webs connecting the 
three branes. We found string webs corresponding
to all the known BPS states, {\em i.e.} the $W$-bosons
and $(2n,1)$ dyons, and were able to explain the
discontinuity in the BPS spectrum (``jumping''
phenomenon) as a simple process of decay of a string 
web into a pair of strings. However, we found
additional string webs which did not correspond to
known BPS states. We argued that although these webs were
classically supersymmetric, they should exhibit some sort
of supersymmetry anomaly analogous to the ``s-rule''
\cite{hw}. 
The issue was subsequently resolved in \cite{mns,dhiz},
where a suitable generalization of the s-rule was given,
which eliminates all the extra states from the BPS 
spectrum. 
On the other hand this raises the question of what these
additional string webs correspond to, given that they 
satisfy the aforementioned fundamental properties.

In a parallel line of development, certain 
{\em non-perturbative} stable non-BPS states have been 
identified in string theory
\cite{sen_nonbps,bg,witten_k}.
In all these examples 
the existence of
the non-perturbative state was
predicted by knowledge of a {\em perturbative}
stable non-BPS state, together with some knowledge of
the strong coupling behavior, {\it e.g.} duality.
A notable class of examples is the $\Omega p$-plane --
D$p$-brane system, in which the lowest mass charged state
is not BPS
\footnote{Another class of 
stable non-BPS states 
corresponds to non-planar webs \cite{bk}.}.
The strong coupling behavior is different
for each $p$, but in every case there should exist
a stable non-BPS state. These have been found so
far for $p=4,5,6,$ and $7$. For $p=7$ the state
corresponds to a string web connecting three mutually
non-local 7-branes, which describe the strong coupling
behavior of the $\Omega 7$--D7 system. The web satisfies 
the two fundamental properties, but is nevertheless
non-BPS, as it violates the ``s-rule''
\cite{sen_nonbps}.

The additional string webs found in 
\cite{bf1} fit nicely into this category of states, 
and therefore seem to correspond to stable non-BPS 
states in $N=2$ $SU(2)$ SYM.
Unlike the previous examples however, these non-BPS
states are not predicted by the perturbative picture.
In this paper we shall extend this result, and find
even more stable non-BPS string webs connecting the 3-brane 
and the two 7-branes.
Furthermore, we will find a second curve of marginal 
stability $\C'_M$ (in addition to the usual one $\C_M$), 
across which some of these webs decay.
This suggests that the moduli space of $N=2$ $SU(2)$
SYM theory is actually 
divided into {\em four} 
regions, each of which has a different stable 
(but not necessarily BPS) particle spectrum.
In the {\em semi-classical} regime, {\it i.e.} outside
both curves of marginal stability, the spectrum includes
states of arbitrary even electric charge, and 
{\em arbitrary magnetic charge} (modulo reducibility of the 
corresponding web).

The paper is organized as follows. In section~2
we review the conditions for a string web to be 
supersymmetric, and the construction of the BPS
states in $N=2$ $SU(2)$ SYM. In section~3 we 
relax the above conditions to incorporate stable,
but not necessarily supersymmetric, string webs,
and solve for the stable particle spectrum.
Section~4 contains our conclusions.

\section{String junctions and the BPS spectrum}
\setcounter{equation}{0}

\noindent Let us first recall how the BPS spectrum
is derived in the 3-brane probe picture. 
Consider a string web lying in the $z$-plane, and 
ending on $N$ $(p,q)$ 7-branes and 
$M$ 3-branes, which are transverse to the plane. 
Denote the charges of the external strings 
by $(p_i,q_i)$, where $i=1,\ldots,N$ for those ending on 
7-branes, and $i=N+1,\ldots,M$ for those ending on 
3-branes \footnote{Note that $p_i$ and $q_i$ need not
be mutually prime. However, if $(p_i,q_i)$ have the same
common divisor ($\geq 2$) for all $i$ the configuration 
is reducible, and therefore at most marginally bound.}. 
The strings are taken to be oriented outward, and
ordered counterclockwise. 
We define the {\em self-intersection} number of the web as
\cite{iqbal}
\be
 \I \equiv 
  \sum_{1\leq i<j\leq N+M}\left|
  \begin{array}{ll}
   p_i & p_j \\
   q_i & q_j
  \end{array} \right|
  - \sum_{i=1}^N\Big(\mbox{gcd}(p_i,q_i)\Big)^2\;.
\ee
The first contribution is due to the string junctions, 
and the second to the strings ending on 7-branes.
To preserve supersymmetry, the webs must satisfy the 
following
\footnote{The complete list of necessary and sufficient 
conditions for supersymmetry in the string picture
is not known. In the lift to M-theory string webs become
membranes with boundaries, in which case a necessary and
sufficient condition for supersymmetry is
that the membranes wrap holomorphic curves.}:
\begin{enumerate}
\item[{\bf a.}] Strings lie on trajectories of 
minimal mass,
{\it i.e.} $(p,q)$-geodesics. 
\item[{\bf b.}] The orientations of the strings are 
correlated with their $(p,q)$ type, such that 
\be
 \theta_{p,q}(z) = \left\{
 \begin{array}{c}
   \arg(p+q\tau(z)) + \theta_{1,0}(z) \\
   \mbox{or}  \\
   \arg(p+q\overline{\tau(z)}) + \theta_{1,0}(z)
 \end{array} \right.
\label{orient}
\ee
These reduce to the zero force condition 
when applied to the junction points.
\item[{\bf c.}] The self-intersection number of the web
satisfies the following inequality \footnote{This follows
from the M-theory picture, since holomorphic curves
embedded in surfaces of vanishing first Chern class 
(like $T^4$ and $K3$) have a self-intersection
number $\# = 2g-2+b$, where $b$ is identified with
the number of co-prime strings not ending on 7-branes.}

\be
 \I \geq -2 + 
    \sum_{i=N+1}^{N+M}
    \mbox{gcd}(p_i,q_i) \; .
\label{Jcondition}
\ee
In fact, this inequality must be satisfied for each
irreducible sub-web, where strings not ending on 7-branes 
count as strings ending on 3-branes.
\end{enumerate}
Conditions {\bf a} and {\bf b} guarantee that the string
web preserves supersymmetry at the {\em classical} level.
At the {\em quantum} level there may be supersymmetry
anomalies, {\it i.e.} the true ground state of the
configuration may break supersymmetry. The so-called
``s-rule'' \cite{hw} is an example of such
an anomaly. For our setup the statement of the s-rule
is that given a $(p,q)$ 7-brane and an $(r,s)$ string
transverse to it, the number of $(p,q)$ strings that can
link the two while still preserving supersymmetry is
bounded above by $|ps-qr|$ \cite{bf2}.
Condition {\bf c} is a generalization of this rule.

The background corresponding to 
pure $SU(2)$ SYM consists of two mutually
non-local 7-branes. Using our previous conventions
\cite{bf1,bf2}, 
the 7-branes are located at $z=+1$ and
$z=-1$, and their charges are $(0,1)$ and $(2,\pm 1)$,
respectively.
Both branch cuts extend along the negative real axis.
The sign for the second 7-brane is $+$ when viewed 
from above the cut and $-$ when viewed from below the cut.
The $(p,q)$-metric on the transverse plane is given 
by \cite{sen_bps}
\be
 T_{p,q}ds = |pda + qda_D| \; ,
\ee
where $a(z)$ and $a_D(z)$ are the integrals of the
Seiberg-Witten differential over the two cycles of
the auxiliary Riemann surface \cite{sw}. For the purpose of 
numerical analysis they are best expressed in terms of
hypergeometric functions,
\begin{eqnarray}
 a(z)& =& \left({z+1\over 2}\right)^{1/2}
     F\left(-\half,\half,1;{2\over z+1}\right)
    \nonumber\\
 a_D(z)&=&  i\left({z-1\over 2}\right)
    F\left(\half,\half,2;{1-z\over 2}\right)\; .
\label{hypergeo}
\end{eqnarray}
Thus $(p,q)$-geodesics satisfy the equation
\be
 p {da\over dt} + q {da_D\over dt} = 
 (p+q\tau(z)){da\over dt} = c_{p,q}  \;,
\label{geodesic}
\ee
and their tangents are therefore oriented along
\be
  \theta_{p,q}(z)
    = \arg(p + q\overline{\tau(z)})
        - \arg(da/dz) + \arg(c_{p,q}) \;.
\label{geodesic_tangent}
\ee
It is immediately clear that the second equation
in (\ref{orient}) is compatible with 
(\ref{geodesic_tangent}),
whereas the first one is not.
This is because the 7-brane background picks
out a unique complex structure on the transverse
plane, in which only strings oriented according
to $p+q\bar{\tau}$ can be supersymmetric \cite{iqbal}.
It then follows that conditions {\bf a} and {\bf b} 
are satisfied 
if and only if
\be
 \arg{c_{p,q}} = \phi \qquad \mbox{for all $(p,q)$ 
strings} 
  \;,
\label{phases1}
\ee
where $\phi$ is arbitrary.

Denote the positions of the $(0,1)$ 7-brane,
$(2,\pm 1)$ 7-brane, and 3-brane by
$z_1, z_2$, and $z_3$, respectively. Likewise,
denote the charges of the strings ending on the
branes by $(p_i,q_i)$, where $i=1,2,3$.
We assume that the string webs are completely
{\em degenerate}, {\it i.e.} that all internal strings
have a vanishing length, so that there is effectively
a single junction point at $z_0$.\footnote{The webs could in 
principle possess hidden faces, but since these correspond
to zero modes, the assumption of degeneration will not alter
our result. As it turns out, none of the solutions 
have zero modes however.}
The geodesic constants are therefore given by
\be
 c_{p_i,q_i} = p_i\Big(a(z_i)-a(z_0)\Big)
      + q_i\Big(a_D(z_i)-a_D(z_0)\Big) \;.
\label{c}
\ee
For the strings ending on 7-branes these simplify since
$p_ia(z_i) + q_ia_D(z_i) = 0$ for $i=1,2$. 

Solutions to (\ref{phases1}) were found for 
$(p_3,q_3)=(\pm 2,0)$ and $(2n,\pm 1)$,
corresponding respectively to the $W$-bosons and  
dyon hypermultiplets.
Furthermore, these webs satisfy condition
{\bf c}, so the corresponding states are indeed BPS 
saturated.
In all cases the solution restricts the junction to 
lie on the
curve defined by
\be
 \mbox{Im}{a_D(z_0)\over a(z_0)}=0 \;, \quad
 \mbox{Re}{a_D(z_0)\over a(z_0)} \in
 \left\{
 \begin{array}{ll}
  [-2,0] & \mbox{if}\;\; z_0\in H^+ \\
  {}[0,+2] & \mbox{if}\;\; z_0\in H^-
 \end{array} \right. \;,
\label{marginal}
\ee
and the 3-brane to lie on a curve defined by
\be 
  \mbox{Im}{p_3a(z_3)+q_3a_D(z_3)\over 
  p_3a(z_0)+q_3a_D(z_0)} = 0 \qquad , \qquad
 {\mbox Re}{p_3a(z_3)+
  q_3a_D(z_3)\over p_3a(z_0)+q_3a_D(z_0)} > 1 \; .
\label{const_phase}
\ee
The first curve is well known as the curve of marginal
stability $\C_M$, and the second is a curve of constant
phase for the function $p_3a(z)+q_3a_D(z)$.
The second condition in (\ref{const_phase}) implies
that the 3-brane must lie outside $\C_M$. 
Thus the string junction picture confirms that,
with the exception of the states $(0,1)$ and $(2,1)$,
which correspond to single strings, the above BPS states 
exist only outside $\C_M$ (figure~1). 
\begin{figure}[htb]
\centerline{\epsfxsize=2.5 in
\epsffile{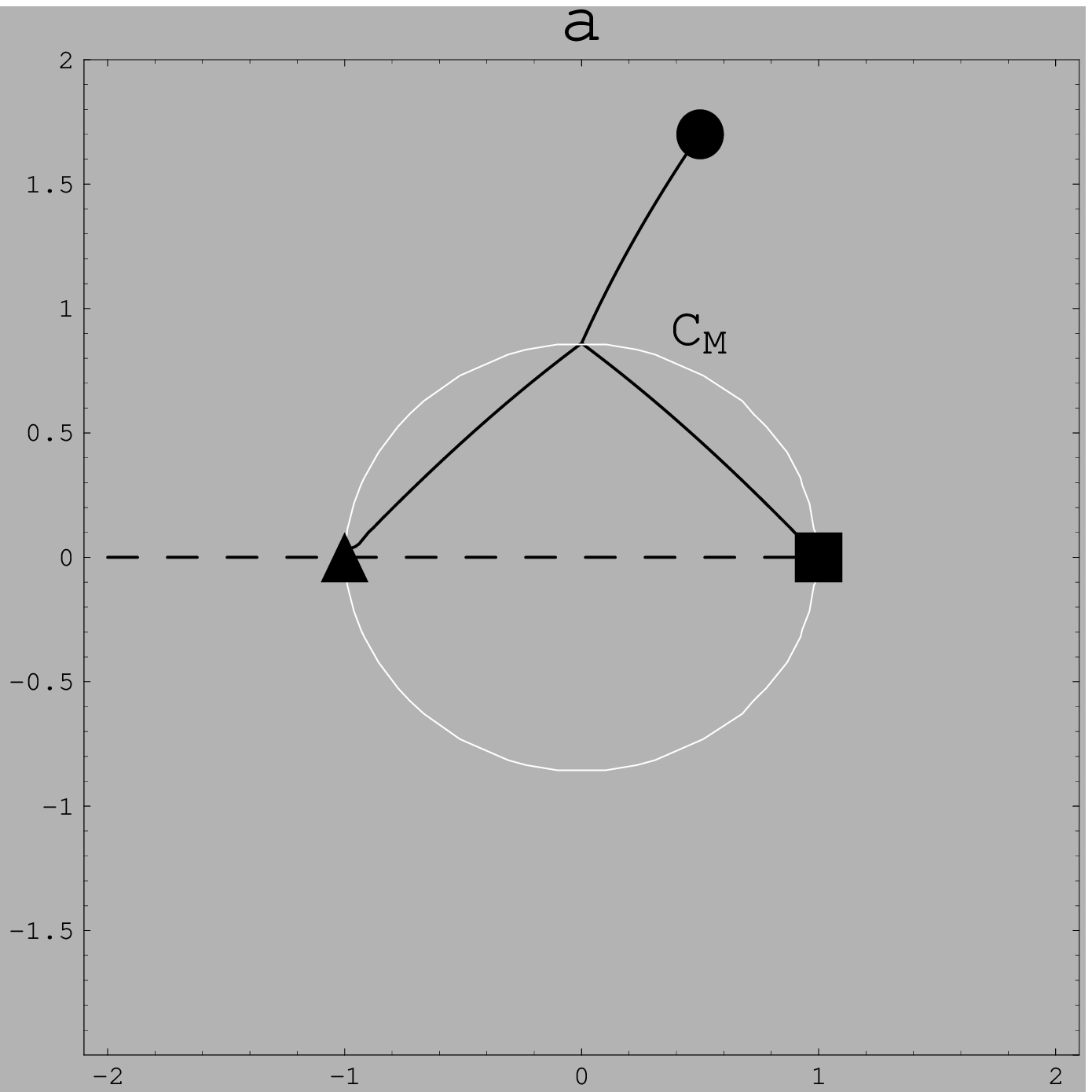}\hspace{1 cm}
\epsfxsize=2.5 in
\epsffile{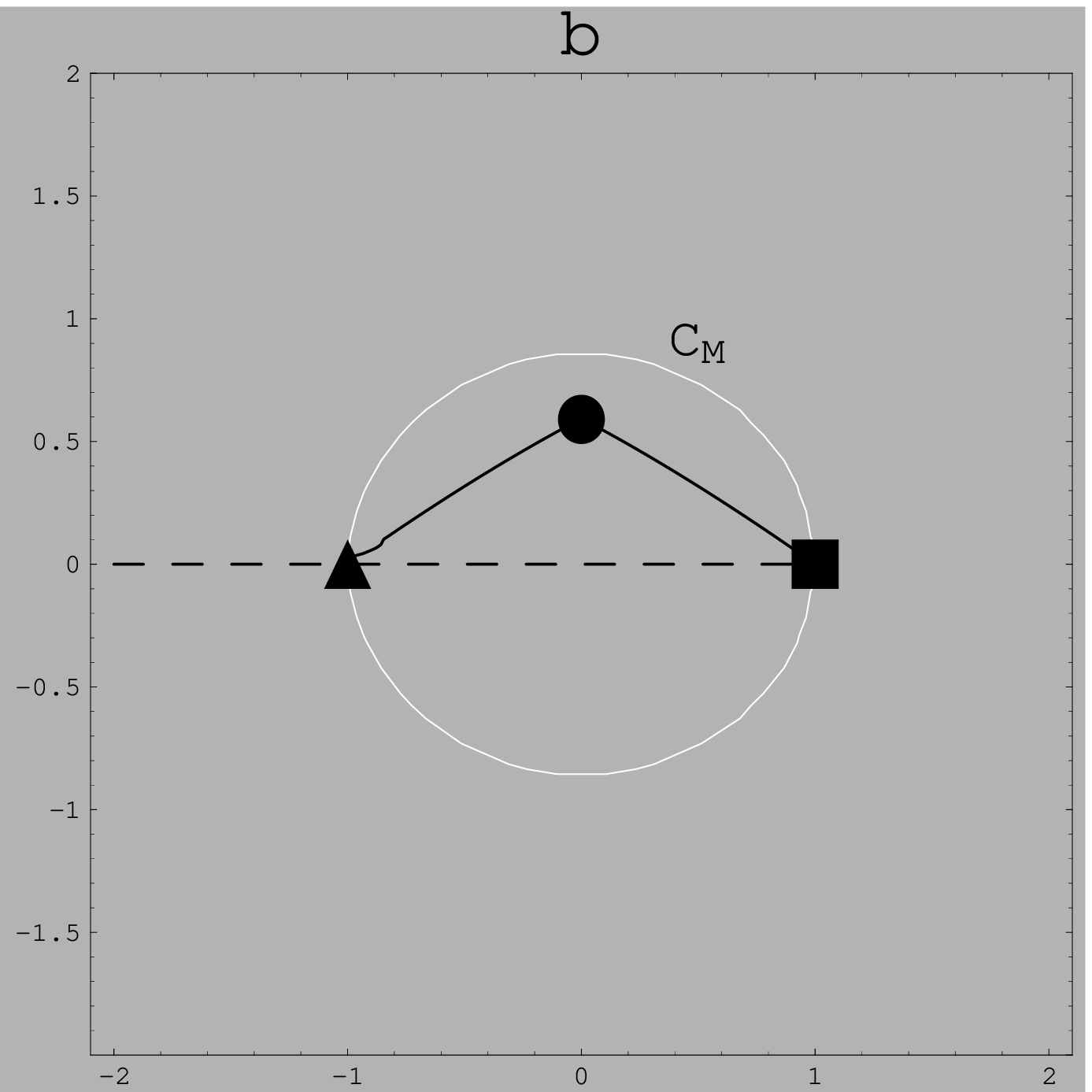}}
\caption{(a) 3-brane outside $\C_M$, BPS state exists.
(b) 3-brane inside $\C_M$, BPS state decays into $(0,1)$'s
and $(2,\pm 1)$'s.}
\end{figure}

\section{String junctions and stable non-BPS states}
\setcounter{equation}{0}

\noindent There are actually additional solutions to
(\ref{phases1}) \cite{bf1}.
However, as these carry magnetic charges of magnitude
greater than $1$, they can easily be shown to violate 
condition {\bf c} \cite{dhiz}.
They do not therefore correspond to BPS states.
On the other hand, the fact that they satisfy
conditions {\bf a} and {\bf b} suggests that 
they are stable string webs, and therefore that we
identify them as stable non-BPS states in the field theory.

More generally, classical stability of a string web
requires only that all the forces cancel.
This means that for stable (but not necessarily
supersymmetric) string webs 
condition {\bf b} is replaced by:
\begin{description}
\item[b'.] The orientations of the strings {\em at the 
junctions} satisfy
\be
 \theta_{p,q}(z_0) = \left\{
 \begin{array}{c}
   \arg(p+q\tau(z_0)) + \theta_{1,0}(z_0) \\
   \mbox{or}  \\
   \arg(p+q\overline{\tau(z_0)}) + \theta_{1,0}(z_0),
 \end{array} \right. 
\label{tension}
\ee
\end{description}
which guarantees that the forces on the junctions due 
to the string tensions cancel. 
%
%
This generalizes condition {\bf b}, which restricted the 
orientations at {\em all} points along the strings.

For string webs which satisfy {\bf b'} but not
{\bf b} supersymmetry is completely broken, and therefore
there are actually non-vanishing forces between the strings.
This means that their trajectories deviate from $(p,q)$-geodesics,
and condition {\bf a} must be modified.
The deviation is such that the string-string
forces are balanced by elastic forces,
\be
 \Delta r/l_s^2 \sim
 l_s^5/ r^6 \;,
\ee 
and is therefore negligible 
in the limit where the distances (in the string metric) between
the branes are much greater than $l_s$.
We can therefore approximate the trajectories by 
$(p,q)$-geodesics in this limit.
Since this is precisely the regime where the stringy picture makes 
sense, we conclude that 
conditions {\bf a} and {\bf b'} guarantee that
the string web is classically stable.
On the other hand, the field theory description holds
when all distances are much {\em smaller} than $l_s$.
Since the above string webs are not supersymmetric, their
existence does not a-priori guarantee that the corresponding
non-BPS states will be stable in the field theory.
We shall return to this point later on.

Both orientations in (\ref{tension}) are now compatible 
with the geodesic condition.
This means there are two kinds of junctions which
can potentially be stable. Let us refer to them 
as $\tau$-junctions and $\bar{\tau}$-junctions,
according to whether we use $(p+q\tau)$ or 
$(p+q\bar{\tau})$ to determine the orientations at the
junction point. These correspond to the two different
ways of ordering the three strings around the junction
(and therefore $\I_\tau = -\I_{\bar{\tau}}$).

For a given position of the 3-brane, a state with a 
given set of charges $(p_3,q_3)$ (other than $(2,\pm 1),
(0,1)$) can either correspond to a $\tau$-junction
or to a $\bar{\tau}$-junction, but not both.
We can understand this qualitatively as follows. 
As $\tau\rightarrow\bar{\tau}$ the ordering of the 
strings changes. 
If we keep the 3-brane fixed,
the $(0,1)$ and $(2,1)$ strings are exchanged.
So for one of the junctions they would have to cross 
in order to end on the appropriate 7-branes. This
leads to a closed loop of string, which is unstable 
to shrink to a point, giving back the other junction
(figure 2).
This will be verified by solving the stability
conditions ${\bf a}$ and ${\bf b'}$, and finding
all the stable states corresponding to $\tau$-junctions
and $\bar{\tau}$-junctions. 
\begin{figure}[htb]
\epsfxsize=5 in
\centerline{\epsffile{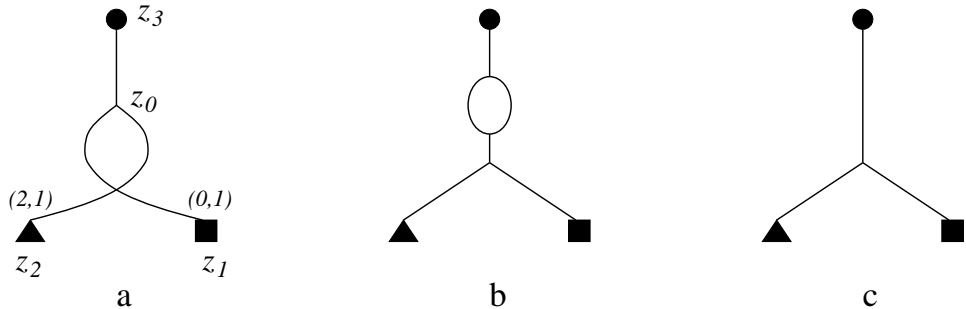}}
\caption{Starting with a $\tau$ (or $\bar{\tau}$) 
junction in which the strings cross (a), the crossing point
splits into two junctions (b), and the loop of string
contracts to a point giving a $\bar{\tau}$ (or $\tau$)
junction (c).}
\end{figure}

As in the previous section, combining conditions
{\bf a} and {\bf b'} we find conditions on the 
arguments of $c_{p_i,q_i}$,
\be
 \arg{c_{p_i,q_i}} = \left\{
 \begin{array}{ll}
 \phi + \arg(p_i + q_i\tau(z_0))^2 & 
            \tau-\mbox{junction} \\
 \phi & \bar{\tau}-\mbox{junction}
 \end{array} \right. \;.
\label{phases2}
\ee
Since $(p_1,q_1)$ and $(p_2,q_2)$ must be integer multiples
of $(0,1)$ and $(2,\pm 1)$, respectively, we can set
$(p_3,q_3)=(2n,m)$. Therefore $(p_1,q_1)=(-m\pm n)(0,1)$,
and $(p_2,q_2)=-n(2,\pm 1)$,
where the sign depends on 
whether the string ending on the 7-brane at $z_1=-1$ comes
from the upper half-plane (upper sign) or the lower half-plane 
(lower sign).
Given (\ref{c}), the conditions in (\ref{phases2}) then imply 
\be
 \mbox{Im}f(z_0) = 0\;,\qquad
 \mbox{sgn}(f(z_0)) = 
 \mbox{sgn}\left({m \mp n\over n}\right) \; ,
\label{marginal2}
\ee
and
\be
 \mbox{Im}\,g(z_0,z_3)=0\;,\qquad
  \mbox{sgn}(g(z_0,z_3)) = 
  \mbox{sgn}(m \mp n) \;,
\label{D3-brane}
\ee
where
\be
 f(z) \equiv {2a(z)+a_D(z)\over a_D(z)} \times
  \left\{
  \begin{array}{ll}
   \tau(z)^2/(2+\tau(z))^2 & \tau-\mbox{junction} \\
   1 & \bar{\tau}-\mbox{junction}
  \end{array}\right. \;,
\ee
and
\be 
 g(z,w) \equiv 
   {2n\Big(a(w)-a(z)\Big) + 
   m\Big(a_D(w)-a_D(z)\Big)\over a_D(z)} \times
 \left\{
 \begin{array}{ll}
   \tau(z)^2/(2+\tau(z))^2 & \tau-\mbox{junc.} \\
   1 & \bar{\tau}-\mbox{junc.}
  \end{array}\right. \;.
\ee
Solutions to (\ref{marginal2}) and (\ref{D3-brane})
will correspond to force-free string webs, and therefore
to stable states. There are four cases to consider,
depending on whether the junction is $\tau$ or $\bar{\tau}$,
and whether the junction point $z_0$ is in the upper
or lower half-plane.
\begin{description}
\item[$\bar{\tau}$-junctions] $\mbox{Im}f=0$ corresponds to
the familiar curve of marginal stability $\C_M$.
On the other hand, it follows from (\ref{marginal}) that
$\mbox{Re}f<0$ on the upper half-plane segment of $\C_M$, 
and $\mbox{Re}f>0$ on the lower half-plane segment. 
Solutions to (\ref{marginal2}) therefore
exist only for $(n-m)/n>0$ and $(n+m)/n>0$ in the upper
and lower half-planes, respectively.
\item[$\tau$-junctions] $\mbox{Im}f=0$ defines
a {\em new} closed curve $\C'_M$ which is diffeomorphic 
to $\C_M$, and, like $\C_M$, intersects the two 7-branes 
(figure 3).
However the behavior of $\mbox{Re}f$ is precisely the
opposite: $\mbox{Re}f>0$ on the upper half-plane segment of 
$\C'_M$, and $\mbox{Re}f<0$ on the lower half-plane segment
(this has been verified numerically).
Consequently solutions to (\ref{marginal2})
exist only for $(n-m)/n<0$ and $(n+m)/n<0$ in the upper
and lower half-planes, respectively.
Therefore, as promised, $\tau$-junctions and
$\bar{\tau}$-junctions are completely complementary.
\end{description}
Once the junction position $z_0$ has been fixed on 
either $\C_M$ or $\C'_M$, the condition $\mbox{Im}g=0$ 
determines a curve intersecting $z_0$ on which the 
3-brane is to be placed. Since $g=0$ when $z_3=z_0$,
{\it i.e.} when the 3-brane is precisely on $\C_M$ 
(or $\C'_M$),
the sign of $g$ determines whether the 3-brane is inside
or outside $\C_M$ (or $\C'_M$). 
The condition on the sign of $g$ in (\ref{D3-brane}) in 
fact
requires the 3-brane to always be {\em outside} $\C_M$ 
(or $\C'_M$) (this has also been verified numerically).
Consequently stable $\bar{\tau}$-junctions can only occur
when the 3-brane is outside $\C_M$, and stable 
$\tau$-junctions can only occur when it is outside $\C'_M$
(figure 3a). When the 3-brane coincides with $\C_M$
($\C'_M$) all $\bar{\tau}$ ($\tau$) junctions become
marginal, and when it moves inside they decay into open 
strings.
Therefore $\C_M$ serves as a curve of marginal stability
for states satisfying $(n-m)/n>0$ in the upper half-plane,
and for states satisfying $(n+m)/n>0$ in the lower 
half-plane (both include the BPS states discussed in
the previous section).
Likewise, the new curve $\C'_M$ is a marginal stability 
curve
for states satisfying $(n-m)/n<0$ in the upper half-plane,
and for states satisfying $(n+m)/n<0$ in the lower
half-plane. 

The picture that emerges is that 
moduli space is divided into four regions (figure 3b). 
In region I only $(0,1)$ and $(2,\pm 1)$ are stable.
In region II only states satisfying
$(n-m)/n>0$ are stable, and in region III
only those satisfying $(n+m)/n>0$ are stable. 
In region IV, which includes the semiclassical regime,
all values of $n$ and $m$ correspond to stable states.
Recall however that if the string web is reducible it can
at most be marginally bound, in which case we will not 
count it as one of the stable states. The condition for
irreducibility is given by
\be
 \mbox{gcd}\Big[\mbox{gcd}(2n,m),n,-m\pm n\Big] = 1 \;,
\ee
where the sign is again correlated with which half-plane
($H^\pm$) we are considering.
\begin{figure}[htb]
\centerline{\epsfxsize=2.5 in
\epsffile{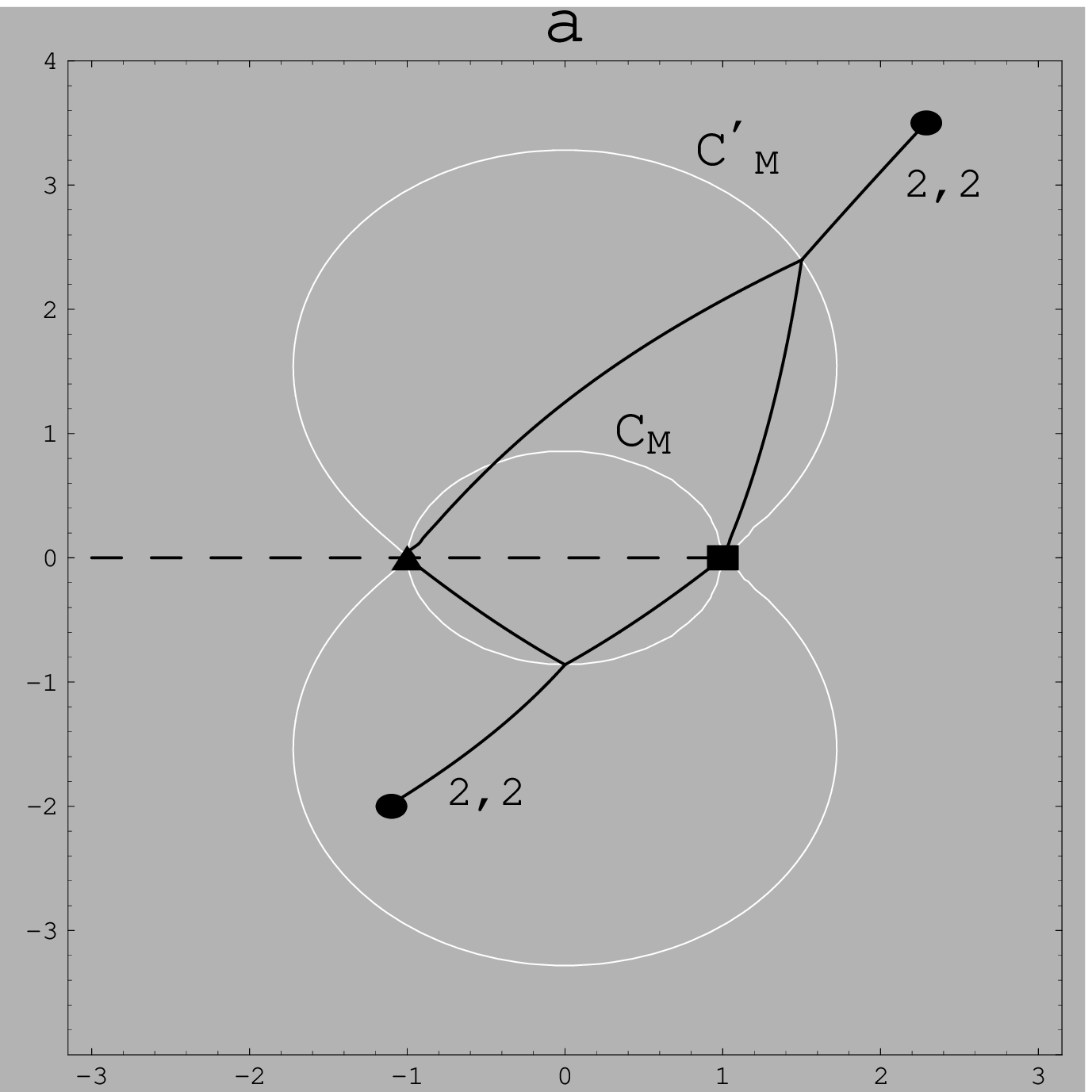}\hspace{1 cm}
\epsfxsize=2.5 in
\epsffile{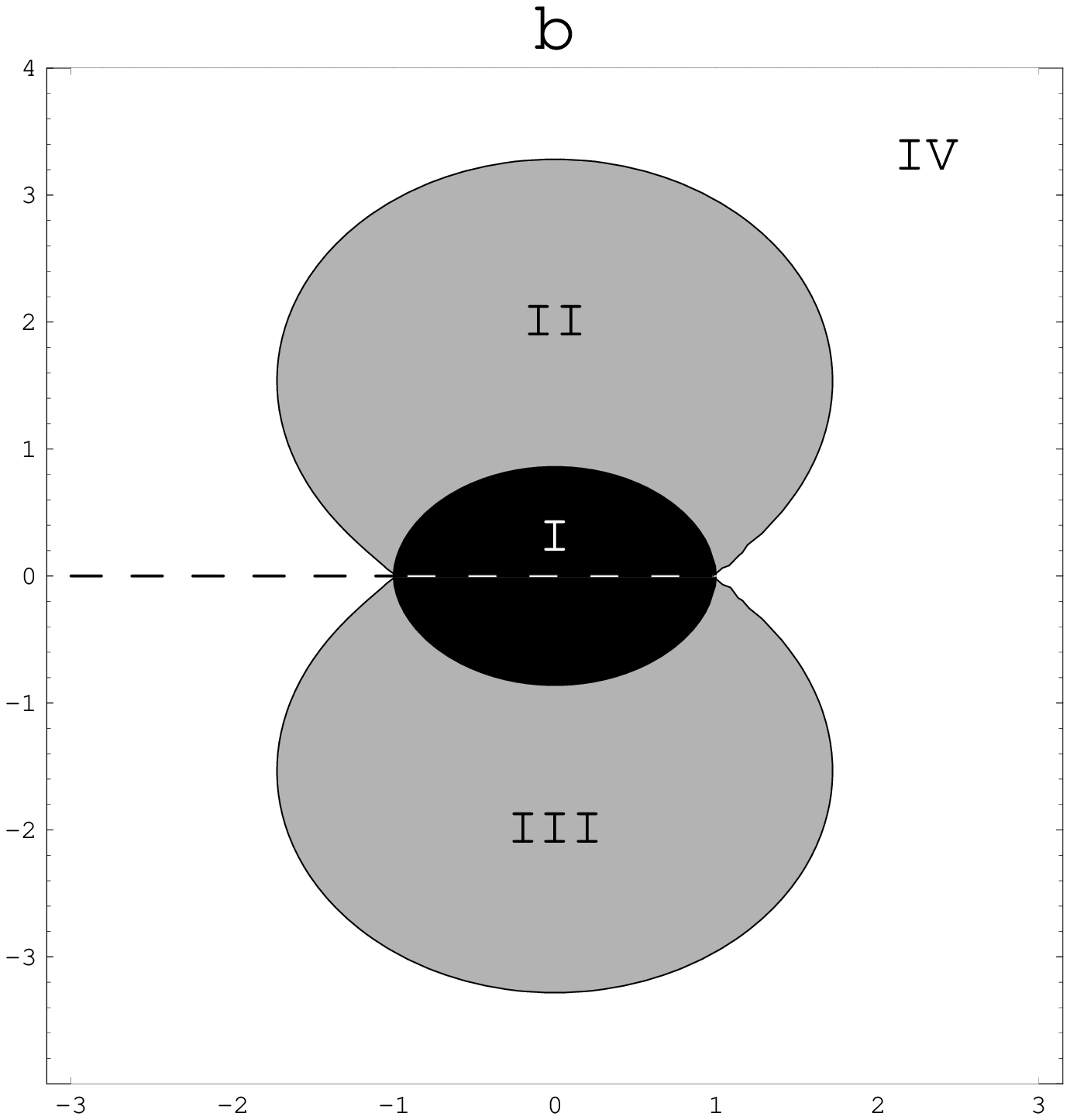}}
\caption{(a) Stable non-BPS string webs; The example
shown has $(2n,m)=(2,2)$, and is therefore given by  
a $\tau$-junction in the upper half-plane, and by a 
$\bar{\tau}$-junction in the lower half-plane.
(b) The two marginal stability curves divide moduli space
into four regions.}
\end{figure}

\medskip

\noindent\underline{$\tau\rightarrow\bar{\tau}$ 
transitions}: As the 3-brane moves around moduli space
(without crossing marginal stability curves) the junction
point slides along the curve of marginal stability, and 
eventually crosses one of the 7-branes, resulting
in a string junction transition, by which strings get 
annihilated or created \cite{bf2}. This does not change
the charges of the string ending on the 3-brane $(2n,m)$, 
since at
the same time the 7-brane at $z_2$ changes its identity
from $(2,1)$ to $(2,-1)$ (or vice versa), which precisely
compensates for the string creation/annihilation effect.
In the BPS case the junction point always remains on $\C_M$.
In the non-BPS case, on the other hand, it is possible
for the junction point to ``jump'' from $\C_M$ to $\C'_M$,
or vice versa. 
This is because the spectrum
in region II is different from that in region III, so
there are some states which correspond to $\tau$-junctions
in one half-plane and to $\bar{\tau}$-junctions in the 
other half-plane, like for example the state $(2,2)$ shown 
in figure~3a.

\section{Conclusions}

\noindent In this paper we have extended the application
of string junctions to the construction of stable non-BPS
states in $N=2$ $SU(2)$ SYM. These correspond to string
webs satisfying a reduced set of properties, whereby
the webs are classically stable, but do not preserve
supersymmetry. 
The non-BPS webs introduce a new curve of marginal
stability into the moduli space, which in addition to the
original one divides moduli space into four regions.
In the region outside both curves we expect
to see stable non-BPS dyons of arbitrary even electric 
charge and arbitrary magnetic charge. The stable particle
spectrum would thus be invariant under $\Gamma(2)$. 
It would be worth while to find these states
directly in the field theory.

It should be stressed that all the results are classical,
and therefore subject to quantum corrections.
In particular, the webs we have constructed exist
in the limit where all distances are much greater than
$l_s$. Since the field theory limit corresponds to all
distances being much smaller than $l_s$, it does not
immediately follow that these webs correspond to stable
non-BPS states in the field theory.
One could however imagine starting with the string web
picture, and gradually reducing the distances between the branes,
while keeping the 3-branes always outside the curves of
marginal stability. In this way the state should remain stable
down to the field theory limit.
The masses of the states and the precise form of the marginal 
stability curves (with regard to the non-BPS states) 
will of-course differ from their classical (string web) values.

\section*{Acknowledgments}

I wish to thank Ansar Fayyazuddin, Tamas Hauer, 
Amer Iqbal, Joe Minahan and Barton Zwiebach
for useful conversations. This work was supported in part
by the DOE under grant no. DE-FG03-92-ER 40701.


\begin{thebibliography}{99}

\bibitem{schwarz} J.H.~Schwarz, 
``Lectures on superstring and $m$ theory dualities: Given at
$ictp$ spring school and at $tasi$ summer school'', 
{\em Nucl. Phys. Proc. Suppl.} {\bf 55B} (1997) 1,
{{\tt hep-th/9607201}}.

\bibitem{sen_net} A.~Sen, 
``String network'',
JHEP 03 (1998) 005, 
{{\tt hep-th/9711130}}.

\bibitem{gz}M.R.~Gaberdiel and B.~Zwiebach,
``Exceptional groups from open strings'',
{\em Nucl. Phys.} {\bf B518} (1998) 151,
{{\tt hep-th/9709013}}.

\bibitem{bergman}
O.~Bergman, ``Three pronged strings and 1/4 {BPS} states in 
$N=4$ super-Yang-Mills theory'', 
{{\tt hep-th/9712211}}.

\bibitem{bk}
O.~Bergman and B.~Kol,
``String webs and 1/4 BPS monopoles'',
{{\tt hep-th/9804160}}.

\bibitem{bf1} 
O.~Bergman and A.~Fayyazuddin,
``String junctions and BPS states in 
Seiberg-Witten theory'',
{{\tt hep-th/9802033}}.

\bibitem{bf2}
O.~Bergman and A.~Fayyazuddin,
``String Junction Transitions in the Moduli Space of N=2 SYM'',
{{\tt hep-th/9806011}}.

\bibitem{mns} 
A.~Mikhailov, N.~Nekrasov, S.~Sethi,
``Geometric realizations of BPS states in N=2 theories'',
{{\tt hep-th/9803142}}.


\bibitem{dhiz}
O.~DeWolfe, T.~Hauer, A.~Iqbal, B.~Zwiebach,
``Constraints on the BPS Spectrum of N=2, D=4
Theories With A-D-E Flavor Symmetry'',
{{\tt hep-th/9805220}}.


\bibitem{biz}
O.~Bergman, A.~Iqbal, B.~Zwiebach,
work in progress.


\bibitem{dm} K. Dasgupta and S. Mukhi,
``BPS nature of 3-string junctions'',
{\em Phys. Lett.} {\bf B423} (1998) 261, 
{{\tt hep-th/9711094}}.

\bibitem{bds} T. Banks, M. R. Douglas, 
N. Seiberg, 
{\em Phys. Lett.} {\bf B388}
(1996) 278, 
{{\tt hep-th/9605199}}.

\bibitem{sen_bps} A. Sen, 
{\em Phys. Rev.} {\bf D55} (1997) 2501, 
{{\tt hep-th/9608005}}.


\bibitem{hw} A. Hanany and E. Witten,
{\em Nucl. Phys.} {\bf B492} (1997) 152,
{{\tt hep-th/9611230}}.

\bibitem{sen_nonbps}
A.~Sen,
``Stable Non-BPS States in String Theory'',
JHEP {\bf 6} (1998) 7,
{{\tt hep-th/9803194}};
``Stable Non-BPS Bound States of BPS D-branes'',
{{\tt hep-th/9805019}};
``Tachyon Condensation on the Brane Antibrane System'',
{{\tt hep-th/9805170}};
``SO(32) Spinors of Type I and Other Solitons on
Brane-Antibrane Pair'',
{{\tt hep-th/9808141}};
``Type I D-particle and its Interactions'',
{{\tt hep-th/9809111}}.

\bibitem{bg}
O.~Bergman and M.R.~Gaberdiel,
``Stable Non-BPS D-Particles'',
{{\tt hep-th/9806155}}.

\bibitem{witten_k}
E.~Witten,
``D-Branes and K-Theory'',
{{\tt hep-th/9810188}}.

\bibitem{iqbal}
A.~Iqbal,
``Self-Intersection Number of BPS Junctions in
Backgrounds of Three and Seven-Branes'',
{{\tt hep-th/9807117}}.






\bibitem{sw} N. Seiberg and E. Witten, 
{\em Nucl. Phys.} {\bf B426} (1994) 19;
{\em Nucl. Phys.} {\bf B431} (1994) 484.


\end{thebibliography}
\end{document}